\documentclass[reprint,superscriptaddress,amsmath,amssymb,aps]{revtex4-1}

\usepackage{graphicx}
\usepackage{dcolumn}
\usepackage{bm}
\usepackage{hyperref}

\usepackage{booktabs}
\usepackage{tikz}
\usetikzlibrary{shapes.geometric,shapes.arrows,decorations.pathmorphing}
\usetikzlibrary{matrix,chains,scopes,positioning,arrows,fit}
\usepackage{circuitikz}
\usepackage{pgfplots}
\pgfplotsset{compat=1.14}
\usepgfplotslibrary{units, groupplots, external, fillbetween}

\usepackage{braket}
\usepackage{makecell}
\usepackage{siunitx}
\usepackage[capitalise]{cleveref}

\begin{document}

\title{Hacking single-photon avalanche detector in quantum key distribution via pulse illumination}

\author{Zhihao Wu}
 \affiliation{Institute for Quantum Information \& State Key Laboratory of High Performance Computing, College of Computer, National University of Defense Technology, Changsha 410073, People's Republic of China}
 
\author{Anqi Huang}%
 \email{angelhuang.hn@gmail.com}
\affiliation{Institute for Quantum Information \& State Key Laboratory of High Performance Computing, College of Computer, National University of Defense Technology, Changsha 410073, People's Republic of China}

\author{Huan Chen}
 \affiliation{College of Liberal Arts and Science, National University of Defense Technology, Changsha 410073, People's Republic of China}
 
\author{Shi-Hai Sun}
 \affiliation{School of Physics and Astronomy, Sun Yat-Sen University, Zhuhai 519082, People's Republic of China}
  
\author{Jiangfang Ding}
 \affiliation{Institute for Quantum Information \& State Key Laboratory of High Performance Computing, College of Computer, National University of Defense Technology, Changsha 410073, People's Republic of China}
 
 \author{Xiaogang Qiang}
 \affiliation{National Innovation Institute of Defense Technology, AMS, Beijing 100071, People's Republic of China}
 \affiliation{Institute for Quantum Information \& State Key Laboratory of High Performance Computing, College of Computer, National University of Defense Technology, Changsha 410073, People's Republic of China}

 \author{Xiang Fu}
 \affiliation{Institute for Quantum Information \& State Key Laboratory of High Performance Computing, College of Computer, National University of Defense Technology, Changsha 410073, People's Republic of China}

 \author{Ping Xu}
 \affiliation{Institute for Quantum Information \& State Key Laboratory of High Performance Computing, College of Computer, National University of Defense Technology, Changsha 410073, People's Republic of China}
 
 \author{Junjie Wu}
 \email{junjiewu@nudt.edu.cn}
 \affiliation{Institute for Quantum Information \& State Key Laboratory of High Performance Computing, College of Computer, National University of Defense Technology, Changsha 410073, People's Republic of China}

\date{\today}

\begin{abstract}
Quantum key distribution (QKD) has been proved to be information-theoretically secure in theory. Unfortunately, the imperfect devices in practice compromise its security. Thus, to improve the security property of practical QKD systems, a commonly used method is to patch the loopholes in the existing QKD systems. However, in this work, {\color{black}we show an adversary's capability of exploiting the imperfection of the patch itself to bypass the patch}. Specifically, we experimentally demonstrate that{\color{black}, in the detector under test,} the patch of photocurrent monitor against the detector blinding attack can be defeated by the pulse illumination attack proposed in this paper. We also analyze the secret key rate under the pulse illumination attack, which theoretically confirmed that Eve can conduct the attack to learn the secret key. This work indicates the importance of inspecting the security loopholes in a detection unit to further understand their impacts on a QKD system. The method of pulse illumination attack can be a general testing item in the security evaluation standard of QKD.
\end{abstract}

\maketitle

\section{Introduction}

Information security is a core of cybersecurity in the digital era. Cryptography provides a vital tool to achieve information security in cyber environment, particularly when untrusted channels are used. Unfortunately, the current widely-used public-key cryptographic infrastructure is threaten by a quantum computer~\cite{shor1997}. To defeat the threat from the quantum world, quantum key distribution (QKD)~\cite{bennett1984} based on the laws of quantum mechanics provides a long-term solution, which has been proved its information-theoretical security. The key generated via QKD can be applied to one-time-pad algorithm, guaranteeing information-theoretically secure communication~\cite{nudt1, nudt2}. Due to plenty of efforts, QKD has developed with a fast pace even to be globalized and commercialized, becoming one of the most mature applications in the field of quantum information~\cite{takesue2007, scheidl2009, sibson17, ding2017, eriksson2019}. Owing to such fast development, standardization of QKD is being considered in the European Telecommunications Standards Institute~(ETSI)~\cite{ISGofETSI}, the International Standard Orgnization~(ISO)~\cite{StandardsNews}, and the International Telecommunication Union~(ITU).

However, a practical QKD system may behave differently from its theoretical model which discloses security loopholes that can be exploited by an eavesdropper, Eve, to learn the secret key, compromising the practical security of QKD systems~{\color{black}\cite{brassard2000,lydersen2010a,xu2010,gerhardt2011,bugge2014,huang2016,sajeed2016,huang2018,huang2019,chistiakov2019,gras2019,huang2018a,xu2020}}. To defeat quantum hacking, an effective countermeasure is to employ an innovative protocol, like measurement-device-independent QKD (MDI QKD)~\cite{lo2012} and twin-field QKD (TF QKD)~\cite{lucamarini2018, curty2019}, to remove the threat from loopholes. However, the high demand of technique and relative low key rate of the innovative protocols limit their application and commercialization. Therefore, to relieve the security threat on the QKD systems in use, it is essential to patch the loopholes in the existing QKD systems~\cite{Koehler2018, Koehler2018a}, most of which employ prepare-and-measure QKD protocol.

As countermeasures, patches, instead of ending the hacking story, lead to a new era of quantum hacking -- inspire quantum attackers to conduct a new round of hacking investigation on the patched system. Patches motivate Eve to discover the {\color{black} imperfection of the patches themselves and the remaining loopholes that the patches fail to close.} Our work {\color{black}proposes a more general blinding attack to bypass the photocurrent monitor}. We discover that{\color{black}, for the detector under test,} bright \emph{pulses} can blind APD intermittently and meanwhile bypass the alarm of photocurrent monitor, although reasonable qubit error rate (QBER) is introduced. We call this method as \emph{pulse illumination attack}.

The pulse illumination attack is a more general type of detector blinding attack than the original one that uses continuous wave~(c.w.) light. In this attack, Eve does not only exploit the mode switch between the linear mode and Geiger mode via light shining, but also makes refined use of the hysteresis current after pulse shining to create a period of full blinding and detection control. The deep investigation on the loopholes strengthens Eve's hacking capability, and thus passes the challenge of protecting the QKD system to countermeasure proposers again. This study emphasizes the significance of further investigating the imperfections of single-photon detector, in order to improve the security property of the standard prepare-and-measure QKD systems that are deployed the most in field~\cite{peev2009, sasaki2011, liao2017}. Moreover, this study contributes a general testing item to the security certification list of QKD standard, which is being drafted in several international organizations, e.g.\ ETSI, ISO, and ITU \cite{ISGofETSI, ETSIwhitepaper,StandardsNews}.

\section{From c.w.\ illumination attack to pulse illumination attack}
\label{sec:theory}

In this section, we briefly review the origin blinding attack, i.e., c.w.\ illumination attack, and then introduce the countermeasure of the photocurrent monitor that is believed to be effective to blinding attacks. Finally, we propose pulse illumination attack as a new form of blinding attack that can bypass the photocurrent monitor.

For a BB84 QKD system, Eve can apply the c.w.\ blinding attack on APDs to eavesdrop the information. In the original c.w.\ blinding attack, Eve injects continuous light to generate a huge photocurrent through APD, which lowers the bias voltage and pulls the APD back to the linear mode that is insensitive to a single photon. Then, she conducts a fake-state attack as follows. She first intercepts and measures each state sent by Alice, and resends a trigger pulse encoded by her measurement result to control Bob's clicks as the same as hers. For more details about the original c.w.\ blinding attack, see \cref{sec:c.w.}.


To patch the loophole exploited by the blinding attack, some QKD systems including our testing object in this work adopt a photocurrent monitor as a countermeasure against the blinding attack~\cite{lydersen2010a, yuan2010, huang2016}. This countermeasure bases on an intuitive assumption that a blinding attack will certainly generate a distinguishable low-frequency photocurrent in the circuit of the detector. The monitor extracts the low-frequency photocurrent as an alarm of the blinding attack. Once the extracted photocurrent reaches an alarming threshold, the blinding attack is considered to be launched. Please note that the extracted photocurrent is named as reported photocurrent in the following text.

However, we find that this countermeasure can not completely hinder the pulse illumination attack. This is because the aforementioned assumption about the photocurrent under a blinding attack does not stand when bright optical pulses are sent to blind an APD. In this attack, a group of blinding pulses accumulatively introduces a high photocurrent. This photocurrent is also able to lower the bias voltage across the APD. As a result, the detector is blinded at that time. After the blinding pulses are gone, the detector is still blinded for a certain period, because the photocurrent gradually reduces due to capacitors in the detector. Thus the detector keeps being blinded until the photocurrent becomes fairly weak. Eve can exploit the	 blinded period to launch the fake-state attack to eavesdrop the information. Theoretically, the length of the blinded period is positively correlated with the energy of the blinding-pulse group.

Unlike the constant high photocurrent introduced by the c.w.\ illumination attack, here the photocurrent varies over time. The photocurrent monitor takes this current as high-frequency noise and ignores most of it. Therefore, Eve can apply the pulse illumination attack to eavesdrop the information without being noticed by the photocurrent monitor.

\section{Experimental demonstration}
\label{sec:experiment}
As a third-party evaluator, we conducted tests about the pulse illumination attack on a APD-based single-photon detector module, provided by an independent party. In the tests, we assume that Eve only knows the public information of the detector as prior knowledge to show a real-life hacking scenario. For the single-photon detector module we tested, the frequency of gate signal is \SI{40}{\mega \hertz}, and the photocurrent monitor inside the module first filters the photocurrent in the circuit of the detector via a lowpass filter to avoid high-frequency noise. The alarming threshold is set as \SI{10}{\micro \ampere} {\color{black}by the independent detector provider}, which is an {\color{black} optimum-performance} value to safely detect c.w.\ illumination blinding attack {\color{black} in practice}. The threshold is far lower than the illegitimate value (\SI{31}{\micro \ampere}) when c.w.\ illumination blinding attack works, as well as leaves a margin to the value of normal working state (\SI{1.4}{\micro \ampere}) to avoid false alarms {\color{black}due to occasionally in-field fluctuation}. The tested detector works at \SI{-50} {\celsius} with dark count rate of $5 \times 10^{-6}$ per gate and 3\% after-pulse probability.

\begin{figure}\centering
	\begin{tikzpicture}[controlpanels/.style={yellow!30!brown!20!,rounded corners,draw=black,thick},
	screen/.style={blue!50!black!60!,draw=black},
	trace/.style={green!60!yellow!40!, ultra thick},
	smallbutton/.style={white,draw=black, thick}]
  
  \def \SGen#1{
	  \fill[black!70!,rounded corners=0.5pt] (-6mm,-4.5mm) rectangle ++(2mm,0.5mm);
	  \fill[black!70!,rounded corners=0.5pt] (6mm,-4.5mm) rectangle ++(-2mm,0.5mm);
	  \fill[green!30!blue!30!,rounded corners=1pt,draw=black](-6.5mm,-4mm) rectangle ++(13mm,8mm);
	  \fill[fill=black!40!,draw=black,rounded corners=1pt](-6.25mm,-3.75mm) rectangle ++(12.5mm,7.5mm);
	  \begin{scope}[samples=150]
	  \fill[black!60!,rounded corners=1pt,draw=black](-6mm,-0.5mm) rectangle (3mm,3mm);
	  \fill[screen] (-5.5mm,0mm) rectangle (2.5mm,2.5mm);
	  \draw [white, xshift=-4.2pt] (-10pt, 3pt) to (4pt, 3pt) to (5pt, 5pt) to (6pt, 3pt) to (10pt, 3pt);
	  \fill[black!70] (4.9mm,1.5mm) circle [radius=0.6mm];
	  \fill[black!30] (5.15mm,1.6mm) circle [radius=0.2mm];
	  \draw[black!60, thick, yshift=-1pt] (3.5mm,0mm) to (6.3mm, 0mm);
	  \foreach \xbutton in {-5mm,-3mm, -1mm, 1mm} {
		  \fill[blue!70] (\xbutton,-1mm) rectangle (\xbutton+1mm, -1.5mm);
		  \fill[black!20] (\xbutton,-2mm) rectangle (\xbutton+1mm, -2.5mm);
	  }
	  \fill[black!20] (4.9mm,-1.5mm) ellipse [x radius=0.6mm, y radius=0.3mm];
	  \fill[left color=black!20,right color=black!70] (4.5mm,4.1mm) rectangle (5.5mm,4.7mm);
	  \fill[top color=black!20,bottom color=black!70] (6.5mm,1mm) rectangle (7.2mm,2mm);
	  \fill[top color=black!20,bottom color=black!70] (6.5mm,-1mm) rectangle (7.2mm,-2mm);
	  \coordinate[label= {[xshift=-1.5mm, yshift=0.5mm]80:\sf \tiny ch1}] (#1_c1) at (7.2mm, 1.5mm);
	  \coordinate[label= {[xshift=-1.5mm, yshift=-0.5mm]-80:\sf \tiny ch2}] (#1_c2) at (7.2mm, -1.5mm);
	  \coordinate [label= above left:\sf \tiny trigger](#1_trg) at (5mm, 4.7mm);
	  \node[yshift=-7mm] {\sf \tiny Waveform generator\par};
	\end{scope}
  }
  \def \SGenLegend#1{
	  \begin{scope}[samples=150, xshift=-16mm]
	  \fill[black!70!,rounded corners=0.5pt] (-6.5mm,-4.5mm) rectangle ++(2mm,0.5mm);
	  \fill[black!70!,rounded corners=0.5pt] (6.5mm,-4.5mm) rectangle ++(-2mm,0.5mm);
	  \fill[green!30!blue!30!,rounded corners=1pt,draw=black](-7mm,-4mm) rectangle ++(14mm,8mm);
	  \fill[fill=black!40!,draw=black,rounded corners=1pt](-6.75mm,-3.75mm) rectangle ++(13.5mm,7.5mm);
	  \fill[black!60!,rounded corners=1pt,draw=black](-6mm,-0.5mm) rectangle (3mm,3mm);
	  \fill[screen] (-5.5mm,0mm) rectangle (2.5mm,2.5mm);
	  \node[white, scale=0.7, xshift=-6pt, yshift=5pt] {\tiny \ttfamily40.000MHZ};
	  \fill[black!70] (4.9mm,1.5mm) circle [radius=0.6mm];
	  \fill[black!30] (5.15mm,1.6mm) circle [radius=0.2mm];
	  \draw[black!60, thick, yshift=-1pt] (3.5mm,0mm) to (6.3mm, 0mm);
	  \foreach \xbutton in {-5mm,-3mm, -1mm, 1mm} {
		  \fill[blue!70] (\xbutton,-1mm) rectangle (\xbutton+1mm, -1.5mm);
		  \fill[black!20] (\xbutton,-2mm) rectangle (\xbutton+1mm, -2.5mm);
	  }
	  \fill[black!20] (4.9mm,-1.5mm) ellipse [x radius=0.6mm, y radius=0.3mm];
	  \fill[left color=black!20,right color=black!70] (4.5mm,4.1mm) rectangle (5.5mm,4.7mm);
	  \fill[top color=black!20,bottom color=black!70] (7.1mm,1mm) rectangle (7.7mm,2mm);
	  \fill[top color=black!20,bottom color=black!70] (7.1mm,-1mm) rectangle (7.7mm,-2mm);
	\end{scope}
  }
  \def\DSG#1{
	  \node  [draw=blue!50!black!50, top color=white,bottom color=blue!20, font=\scriptsize \sf,minimum width=12mm, minimum height=5mm, align=center,label={[align=center,font=\tiny \sf ]below:Digital signal \\ generator}] (#1) {DSG};
  
  }
  \def\DSGLegend#1#2{
	  \node  [xshift=-16mm, draw=blue!50!black!50, top color=white,bottom color=blue!20, font=\sf,minimum width=12mm, minimum height=5mm] (#2) {\scriptsize #1}; 
  }
  
  \def\Attn#1#2{
	  \node  [draw=black!50, top color=white,bottom color=black!20, font=\sf,minimum width=12mm, minimum height=5mm] (#2) {\scriptsize #1}; 
  }
  \def\SA#1{
	  \node  [draw=black!50, top color=white,bottom color=black!20, font=\scriptsize \sf,minimum width=10mm, minimum height=5mm, text width=10mm,align=center,label={[align=center,font=\tiny \sf ]below:Manual variable \\ attenuator}] (#1) {MVA};
  }
  \def\VOA#1{
	  \node  [draw=black!50, top color=white,bottom color=black!20, font=\scriptsize \sf,minimum width=10mm, minimum height=5mm, text width=10mm,align=center,label={[align=center,font=\tiny \sf ]below:Digital variable \\ attenuator}] (#1) {DVA};
  }
  \def\SALegend{
	  \node  [draw=black!50, top color=white,bottom color=black!20, font=\sf,minimum width=12mm, minimum height=5mm]  {\scriptsize SA}; 
	  \node [xshift=12mm,align=left, inner sep=0pt, text width=10mm, minimum height=5mm, font=\tiny \sf]{Screw attenuator};
  }
  \def\VOALegend{
	  \node  [draw=black!50, top color=white,bottom color=black!20, font=\sf,minimum width=12mm, minimum height=5mm]  {\scriptsize VOA}; 
  \node [xshift=12mm, align=left, inner sep=0pt, text width=10mm, minimum height=5mm, font=\tiny \sf]{Digital variable attenuator};
  }
  \def\Bsignal#1{
	  \node[minimum width=12mm, label=below:\tiny \sf Blinding signal] (#1) {};
	  \foreach \pulse in {-6mm,-3mm,-0mm, 3mm}
		   \draw[blue!70!black]  (\pulse,0mm) -| ({\pulse+1mm}, 3mm) -| ({\pulse+2mm}, 0mm) -- ({\pulse+3mm}, 0mm) ; 
  }
  \def\Tsignal#1{
	  \node[minimum width=12mm, label=below:\tiny \sf Trigger signal] (#1) {};
	  \begin{scope}[blue!70!black]
	  \draw  (-6mm,0mm) -- (-0.5mm, 0mm) ; 
	  \draw  (0.5mm,0mm) -- (6mm, 0mm) ; 
	  \draw  (-1.5mm,0mm) -| (-0.5mm, 1mm) -| (0.5mm, 0mm) -- ({1.5mm+1mm}, 0mm) ; 
	  \end{scope}
  }
  
  \def\Laser#1{
	  \node[draw=red!50!black!50,top color=white, bottom color=red!50!black!20,minimum width=12mm, minimum height=5mm, label={[font=\sf \tiny]below:\SI{1550}{\nano \meter}}] (#1) {};
	  \begin{scope}[line cap=round, very thick, red!80]
		  \draw(-3mm,-1.5mm) to (-3mm, 1.5mm) ;
		  \draw[rotate around={45:(-3mm,0mm)}] (-3mm,-1.5mm) to (-3mm, 1.5mm) ;
		  \draw[rotate around={-45:(-3mm,0mm)}] (-3mm,-1.5mm) to (-3mm, 1.5mm) ;
		  \draw(-4.5mm, 0mm) to (4.5mm, 0mm) ; 
	  \end{scope}
  } 
  \def\LaserLegend#1{
	  \node[label=right:\tiny \sf 1550nm Laser, draw=red!50!black!50,top color=white, bottom color=red!50!black!20,minimum width=12mm, minimum height=5mm] (#1) {};
	  \begin{scope}[line cap=round, very thick, red!80]
		  \draw(-3mm,-1.5mm) to (-3mm, 1.5mm) ;
		  \draw[rotate around={45:(-3mm,0mm)}] (-3mm,-1.5mm) to (-3mm, 1.5mm) ;
		  \draw[rotate around={-45:(-3mm,0mm)}] (-3mm,-1.5mm) to (-3mm, 1.5mm) ;
		  \draw(-4.5mm, 0mm) to (4.5mm, 0mm) ; 
	  \end{scope}
  } 
   \def\BS#1{
	  \coordinate (#1_west) at (-2mm, 0);
	  \coordinate (#1_east) at (2mm, 0);
	  \node at (0,-2mm) {\tiny \sf 50:50};
	  \shade [top color=white,bottom color=gray] (-2mm,-0.5mm) rectangle +(4mm, 1mm);
	   } 
   \def\BSLegend#1{
	  \coordinate (#1_west) at (-2mm, 0);
	  \coordinate (#1_east) at (2mm, 0);
	  \node at (0,-2mm) {\tiny \sf 50:50};
	  \node[minimum width=12mm, minimum height=5mm, label=right:\tiny \sf Beam splitter] (#1) {};
	  \shade [top color=white,bottom color=gray] (-2mm,-0.5mm) rectangle +(4mm, 1mm);
	  \draw  [-, red!65, very thick] (-2mm,0mm) to [out=180, in=-60] (-4mm, 2mm);
	  \draw  [-, red!65, very thick] (-2mm,0mm) to [out=180, in=60] (-4mm, -2mm);
	  \draw  [-, red!65, very thick] (2mm,0mm) to [out=0, in=240] (4mm, 2mm);
	  \draw  [-, red!65, very thick] (2mm,0mm) to [out=0, in=120] (4mm, -2mm);
	   } 
  \def\APD#1{
	  \node[draw=black, fill=white, font=\sf,minimum width=9mm, minimum height=5mm, label=86:\tiny \sf clk] (#1) {\scriptsize APD}; 
	  \fill[black] (-4.5mm,-1.5mm) arc [start angle=-90, end angle=90, radius=1.5mm] -- cycle;
  }
  \def\APDLegend#1{
	  \filldraw  [black, xshift=0mm] (-9mm,-2mm) arc [start angle=-90, end angle=90, radius=2mm] -- cycle;
	  \node [inner sep=0pt,text width=15mm, minimum height=5mm, font=\tiny \sf]at(2mm,0){APD's active area};
  }
  \def\PM#1{
	  \coordinate (#1_west) at (-4.5mm, 0)  ;
	  \coordinate [label={[align=center,font=\tiny \sf ]right:Power \\ meter}](#1_east) at (-3mm, 0);
	  \draw  [black, fill=white, label=right:\small PM] (-4.5mm,-2mm) arc [start angle=-90, end angle=90, radius=2mm] -- cycle;
  }
  \def\PMLegend#1{
	  \draw  [black, fill=white, xshift=0mm] (-9mm,-2mm) arc [start angle=-90, end angle=90, radius=2mm] -- cycle;
	  \node [align=left, text width=15mm, inner sep=1pt,minimum height=5mm, font=\tiny \sf]at(2mm,0){Power meter};
  }
  
  \matrix (m) at (0, 0) [
	  column sep=2mm,
	  row sep={6 mm,between origins}
	  ]
	{
	& & \DSG{dsg}& & \coordinate (tmp_relay) at (0, -0.5mm); & & \\[6 mm]
	  & \Bsignal{bsignal} & \Laser{blind} & \SA{sa}& &\APD{apd}&\\
	  \SGen{sg}& & & & \BS{bs} & &\\
	  & \Tsignal{tsignal} & \Laser{trigger} & \VOA{voa}& & \PM{pm}&\\
	};
  \begin{scope}[-, blue!50, very thick]
  \draw (bsignal) to  (blind);
  \draw  (tsignal) to  (trigger);
  \draw  (sg_c1) to  [out=0,in=180]  (bsignal);
  \draw  (sg_c2) to  [out=0,in=180]  (tsignal);
  \draw  (dsg) to  [out=0,in=175]  (tmp_relay);
  \draw (tmp_relay) to [out=-5,in=90] (apd);
  \draw  (dsg) to  [out=180,in=90]  (sg_trg);
  \end{scope}
  \begin{scope}[-, red!65, very thick]
  \draw  (blind) to  (sa);
  \draw (trigger) to  (voa);
	  \draw (sa) to  [out=0,in=180] (bs_west);
	  \draw (voa) to  [out=0,in=180] (bs_west);
	  \draw (bs_east) to  [out=0,in=180] (apd);
	  \draw (bs_east) to  [out=0,in=180] (pm_west);
	  \end{scope}
	  \begin{scope}
	  \end{scope}
	  \end{tikzpicture}
	\caption{Experimental setup for the pulse illumination attack. The double-channel arbitrary waveform generator excites \SI{1550}{\nano \meter} lasers to generate the blinding pulses and the trigger pulses. The trigger pulses do not contribute to blind the APD and they are just used for calibrating the blinded period and controlling Bob's click. The manual variable attenuator and the digital variable attenuator modulate the energy of the blinding pulses and trigger pulses precisely. The 50:50 beam splitter merges the blinding pulses and the trigger pulses. We use a digital signal generator to synchronize our blinding pulses and the trigger pulses with the single-photon detector's clock. The power meter monitors the total energy of pulses going to the single-photon detector.}
	\label{fig:bldsetup}
\end{figure}
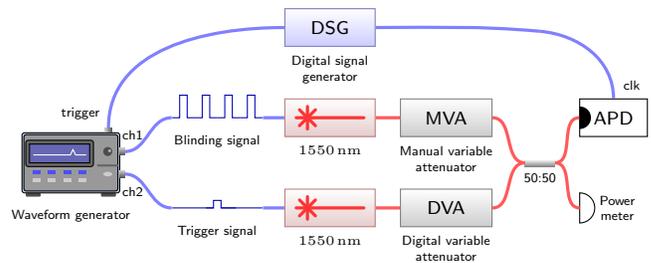

\begin{figure*}
	\centering
	\includegraphics[scale=0.6]{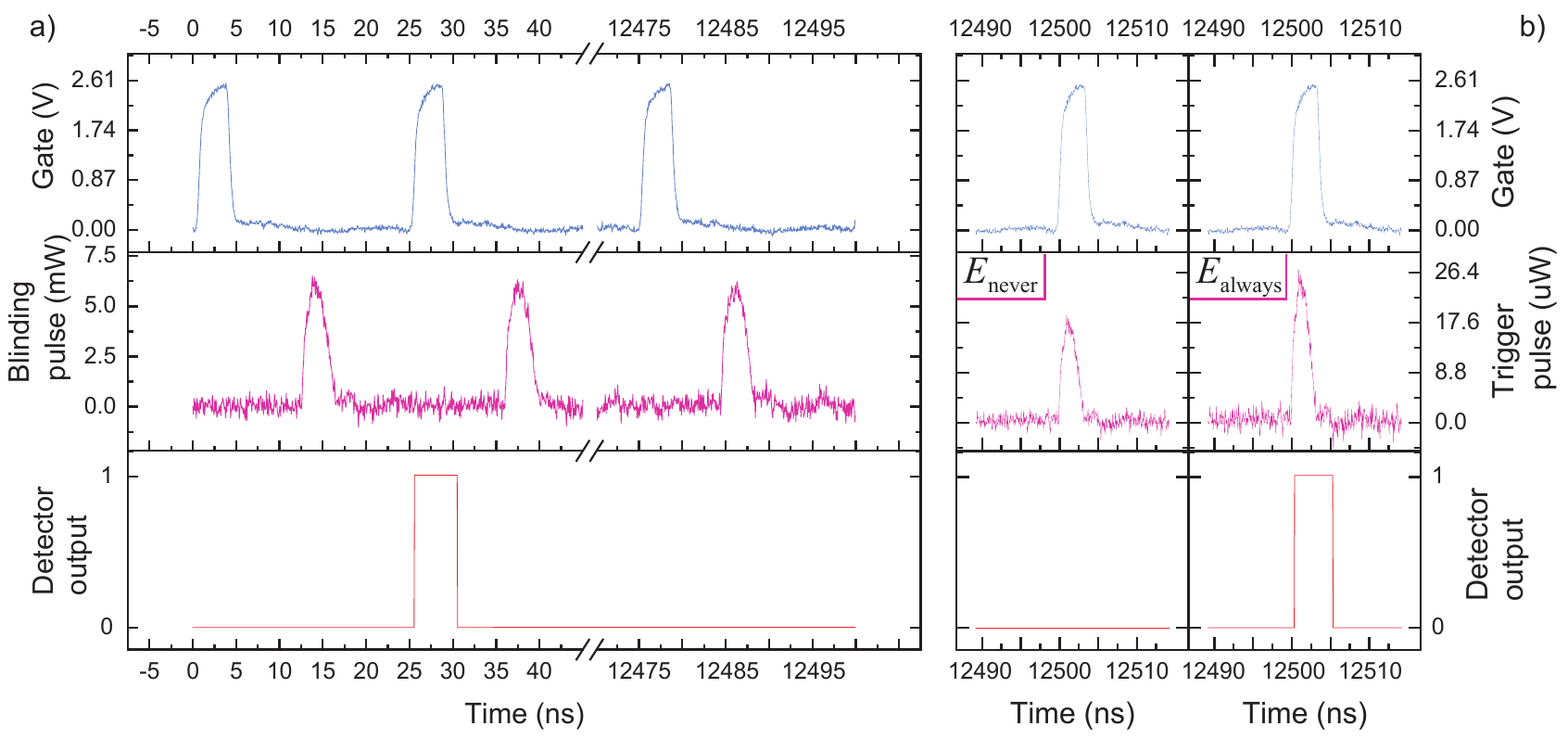}
	\caption{Oscillograms where a) the detector is blinded by a group of 500-cycle blinding pulses and b) a trigger pulse with energy $E_\text{never}$ / $E_\text{always}$ is sent during the blinded period to get no click/click.}
	\label{fig:attack}
\end{figure*}
	
Our experiment setup is shown in \cref{fig:bldsetup}. A digital signal generator synchronize the whole system. The channel 1 of the waveform generator excites a \SI{1550}{\nano \meter} laser diode to launch a group of blinding pulses with the frequency as the same as that of the SPAD gate, \SI{40}{\mega \hertz}. In our experiments, the width of each blinding pulse was set as \SI{4}{\nano \second}, and we kept the energy of each blinding pulse being \SI{13.32}{\pico \joule}. The blinding pulses were applied outside the gate signals to avoid unwanted clicks caused by the blinding pulses. A group of blinding pulses only triggers a click at the beginning of the group, which is followed by \SI{5}{\micro \second} dead time. After that, the detector is blinded due to the accumulated photocurrent and Eve can launch the fake-state attack in this blinded period during which no dark count exists. Long intervals between groups are necessary for reducing the low-frequency photocurrent and avoiding being noticed by the photocurrent monitor. \SI{2}{\milli \second} interval can satisfy such requirement in our testing. This experimental result is shown in \cref{fig:attack}a. The detailed analysis about the parameters of blinding pulses are given in \cref{sec:detail}.

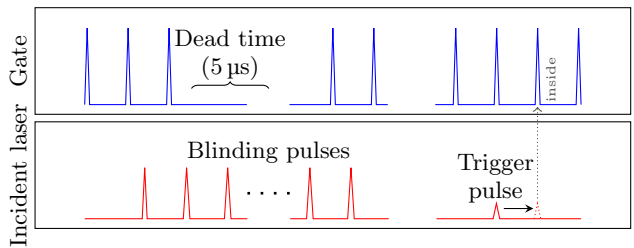
\begin{figure}
	\centering
	  \begin{tikzpicture}
	\begin{groupplot}[
	group style={
	group size=1 by 2,
	x descriptions at=edge bottom,
	vertical sep=0.1cm
	},
	width=9.5cm,
	height=3cm,
	ymax=1.9,
	xtick=\empty,
	ytick=\empty
	]
	
	\nextgroupplot [ylabel=Gate]
	\addplot[blue] coordinates {(0,0) (1.5,1.5) (3,0) (25,0) (26.5,1.5) (28,0) (50,0) (51.5,1.5) (53,0) (99,0)
	
	(125,0) (150,0) (151.5,1.5) (153,0) (175,0) (176.5,1.5) (178,0) (185,0)
	
	 
	 (214,0) (225,0) (226.5,1.5) (228,0) (250,0) (251.5,1.5) (253,0) (275,0) (276.5,1.5) (278,0) (300,0) (301.5,1.5) (303,0)};
	 \nextgroupplot [ylabel=Incident laser, clip=false]
	 \addplot[red] coordinates {(0,0) (35,0) (36.5,1) (38,0) (60,0) (62,1) (64,0) (85,0) (87,1) (89,0) (99,0) 
	
	(125,0) (135,0) (137,1) (139,0) (160,0) (162,1) (164,0) (185,0)
	 
	 (214,0) (248.5,0) (250.5,0.3) (252.5,0) (302,0)};
	 \draw [loosely dotted, line width=1pt] (axis cs:99,0.5) -- (axis cs:125,0.5);
	 \node [] at (axis cs:112,1.3) {Blinding pulses};
	 \node [align=center] at (axis cs:250,0.8) {Trigger \\ pulse};
	 \draw [->,>=stealth] (axis cs:255,0.2) -- (axis cs:273,0.2);
	 \addplot[red, densely dotted] coordinates{(273.5,0) (275.5,0.3) (277.5,0)};
	 \coordinate (bld illm) at (axis cs:137, 1.1);
	 \coordinate (mid bld) at (axis cs:137, 1.5);
	 \coordinate (trg pls) at (axis cs:275.5,0.4);
	 \coordinate (mid trg) at (axis cs:275.5, 1.5);
	 \draw [black!70, densely dotted, ->] (trg pls) to (mid trg) to node [auto,below,font=\tiny,sloped,align=center,xshift=6mm] {inside} +(0, 0.47cm);
	 \draw [decorate, decoration={brace}, yshift=1.8cm] (axis cs: 65, -0.1) -- node [above, align=center] {Dead time \\ (\SI{5}{\micro \second})}(axis cs: 112, -0.1);
	\end{groupplot}
	\end{tikzpicture}
	
	  \caption{The methodology of calibrating the blinded period after a group of blinding pulses. We apply a weak trigger pulse that contains 67 photons as a discriminator of blinding inside the gate after the blinding pulses. We move the trigger pulse to right gate-by-gate and repeat the calibration to probe the boundary of the blinded period.}
	  \label{fig:calibration}
\end{figure}

\begin{table}[htbp]
	\centering
	\caption{The blinded period, the number of fully controllable gates, and the reported photocurrent under pulse illumination with different cycle numbers. The dead time caused by the initial blinding pulse is not included in the blinded period. The interval length between two groups of blinding pulses is set as \SI{2}{\milli \second}. In all these cases, the reported photocurrent is close to that in normal working state, \SI{1.4}{\micro \ampere}. The built-in alarming threshold of the photocurrent monitor is \SI{10}{\micro \ampere}.}
	\label{tab:blinded range}
	\begin{tabular}{p{1.5cm}<{\centering}p{1.5cm}<{\centering}p{1.7cm}<{\centering}p{1.9cm}<{\centering}}
	\toprule
	\makecell[c]{Cycle \\ number} & \makecell[c]{Blinded \\ period \\ (\si{\micro \second})} & \makecell[c]{Fully \\ controllable \\ gates} & \makecell[c]{Reported \\ photocurrent \\ (\si{\micro \ampere})} \\
	\midrule
	250& 2.025& No data& 1.8\\
	300& 20.025& No data& 1.8\\
	350& 45.025& 72& 1.9\\
	400& 100.05& 150& 1.9\\
	450& 135.05& 330& 2\\
	500& 195.05& 690& 2.1\\
	\bottomrule
	\end{tabular}
\end{table}
	
The channel 2 of the waveform generator excites another \SI{1550}{\nano \meter} laser diode to launch a trigger pulse to calibrate the length of the blinded period and the fully controllable range inside. The methodology of calibrating a blinded period is shown in \cref{fig:calibration}. A trigger pulse contains 67 photons, which can trigger a click in Geiger mode but is not strong enough to trigger a click in the linear mode. We first apply the trigger pulse at the gate just after the group of blinding pulses. If the trigger pulse causes no click, the detector is blinded during this gate. The trigger pulse is then moved away from the group of blinding pulses gate-by-gate to repeat the calibration process until the trigger pulse causes a click. The period of no click is the blinded period. The length of the blinded period generated by a group of 250-/300-/350-/400-/450-/500-cycle blinding pulses is shown in \cref{tab:blinded range}. As the trigger pulse contains multiple photons, the length of the blinded period here is just conservative estimations.

\begin{figure}
	\begin{center}
	\begin{tikzpicture}[trim left=-0.5cm]
	\begin{groupplot}[
	group style={
	group size=2 by 2,
	xlabels at=edge bottom,
	ylabels at=edge left,
	},
	height=4cm,
	width=5cm,
	xlabel=Time,
	ylabel=Energy,
	x unit=\si{\micro \second},
	y SI prefix=femto, y unit=\si{\joule}
	]
	
	\nextgroupplot[title=350-cycle, xmin=7.25, xmax=8.75, xtick={7.25, 8.75}]
	
	\addplot+ [smooth,tension=0.2] table[col sep=comma, x=t, y=Ea] {data/350cycle-ahn-zoomin.csv};
	\addplot+ [smooth,tension=0.2] table[col sep=comma, x=t, y=Eh] {data/350cycle-ahn-zoomin.csv};
	\addplot+ [smooth,tension=0.2, mark=triangle*] table[col sep=comma, x=t, y=En] {data/350cycle-ahn-zoomin.csv};
	
	\nextgroupplot[title=400-cycle, xmin=7.25, xmax=11, xtick={7.25,10, 11}]
	\addplot+ [smooth,tension=0.2] table[col sep=comma, x=t, y=Ea] {data/400cycle-ahn-zoomin.csv};
	\addplot+ [smooth,tension=0.2] table[col sep=comma, x=t, y=Eh] {data/400cycle-ahn-zoomin.csv};
	\addplot+ [smooth,tension=0.2, mark=triangle*] table[col sep=comma, x=t, y=En] {data/400cycle-ahn-zoomin.csv};
	
	\nextgroupplot[title=450-cycle, xmin=7.25, xmax=15.5, xtick={7.25,11.25, 15.5}]
	\addplot+ [smooth,tension=0.2] table[col sep=comma, x=t, y=Ea] {data/450cycle-ahn-zoomin.csv};
	\addplot+ [smooth,tension=0.2] table[col sep=comma, x=t, y=Eh] {data/450cycle-ahn-zoomin.csv};
	\addplot+ [smooth,tension=0.2, mark=triangle*] table[col sep=comma, x=t, y=En] {data/450cycle-ahn-zoomin.csv};
	
	\nextgroupplot[title=500-cycle, xmin=7.25, xmax=24.5, xtick={7.25,12.5, 24.5}, legend columns=3, legend entries={$E_{\text{always}}$,$E_{\frac{1}{2}}$,$E_{\text{never}}$}, legend style={at={(-0.1, -0.6)}, anchor={south}}]
	\addplot+ [smooth,tension=0.2] table[col sep=comma, x=t, y=Ea] {data/500cycle-ahn-zoomin.csv};
	\addplot+ [smooth,tension=0.2] table[col sep=comma, x=t, y=Eh] {data/500cycle-ahn-zoomin.csv};
	\addplot+ [smooth,tension=0.2, mark=triangle*] table[col sep=comma, x=t, y=En] {data/500cycle-ahn-zoomin.csv};
	
	\end{groupplot}
	\end{tikzpicture}
	\end{center}
	
	  \caption{$E_{\text{always}}$, $E_{\frac{1}{2}}$ and $E_{\text{never}}$ inside the fully controllable range of a blinded period generated by a group of 350/400/450/500-cycle blinding pulses. {\color{black} The data not satisfying the fully controllable condition ($E_{\text{always}} < 2 E_{\text{never}}$) are not included in this figure. }The time origin is the arriving of the first blinding pulse in the group. The inflection points are the moments that the blinding pulses end, where the accumulated photocurrent reaches to the maximum.}
	  \label{fig:controllablearea}
\end{figure}
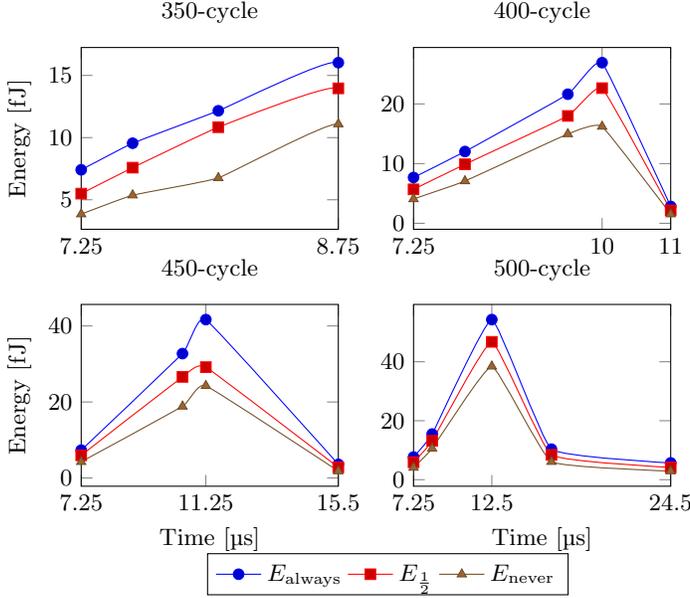
	
By a similar methodology but varying the energy of the trigger pulse, we can further calibrate a fully controllable range. As shown in \cref{fig:attack}b, at each gate inside the blinded period, we vary the energy of a trigger pulse to observe the click probability and record the energy that can trigger a click with the probability of 100\%/50\%/0\% as $E_{\text{always}}$/$E_{\frac{1}{2}}$/$E_{\text{never}}$. If $E_{\text{always}} < 2 E_{\text{never}}$, the detector at this gate in the blinded period is fully controllable by Eve for a BB84 QKD system (while the rest gates of the blinded period also do not have dark counts and are partly controllable). The experimental data of the fully controllable range in the blinded period generated by a group of 350-/400-/450-/500-cycle blinding pulses are shown in \cref{fig:controllablearea}. Note that in all the testing above, the reported photocurrent keeps being far lower than the built-in alarming threshold of the photocurrent monitor (\SI{10}{\micro \ampere}), as shown in \cref{tab:blinded range}.

\section{Security analysis for a decoy-state BB84 QKD system}
\label{sec:simulation}
In this section, we analyse Eve's maximum-profit strategy of attacking a real-life decoy-state BB84 QKD system via pulse illumination, and we further study the threat of this attack to the system. Here the detection parameters are from the Gobby-Yuan-Shields (GYS) experiment~\cite{gobby2004}. The interval length, the blinded period, and the fully controllable range are from our experimental results as shown in \cref{tab:blinded range}.

\subsection{Eve's maximum-profit strategy}
The strategy of Eve's attack is as follows. We assume Eve uses lossless channels to connect Alice and Bob. Without introducing deviation to the normal value of total gain, she launches fake-state attack during the fully controllable range, while blocks or passes the state from Alice during the unblinded time. Therefore, to eavesdrop the maximum information, she needs to optimize the parameters of her attack.

Specifically, three parts constitute the total gain under Eve's attack -- the click trigger by each group's first blinding pulse, the gain under fake-state attack ($Q_\omega^\text{Eve}$) during the fully controllable range, and the gain that Eve blocks or passes the state from Alice during the unblinded period. Regarding the fake-state attack, Eve first measures Alice's state with a perfect detector and resends to Bob by a trigger pulse with energy $E_\text{always}$. Bob's basis choice matches to Eve's half of the time, which triggers a click at Bob's detector with 100\% probability. Thus, $Q_{\omega}^\text{Eve} = \frac{1}{2} (1-e^{-\omega})$, where $\omega \in \{\mu = 0.6, \nu = 0.2, 0\}$ is the mean photon number of the signal state, the decoy state, and the vacuum state. When Eve blocks Alice's states, only dark counts, $Y_0 = 1.7 \times 10^{-6}$, happen at Bob's detector; {\color{black} when Eve passes Alice states via the lossless channel, the corresponding gain is $Q_\omega^\text{pass} = Y_0+1-e^{- \eta_\text{ch} \eta_\text{Bob} \omega} = Y_0+1-e^{- \eta_\text{Bob} \omega}$ (where $\eta_\text{ch} = 1$ is the transmittance efficiency of Eve's lossless channel and $\eta_\text{Bob}$ is the transmittance of Bob's optical device).} Therefore, the total gain under the pulse illumination attack is
{\color{black}
\begin{equation} \label{Q}
\begin{aligned}
Q_\omega &= \frac{1 + p Q_\omega^\text{Eve} N_{\text{control}}}{N_{\text{interval}}} \\
    &+ \frac{(N_{\text{interval}} - N_{\text{blind}} - N_{\text{dead}}) [\gamma Q_{\omega}^{\text{pass}} + (1 - \gamma) Y_0]}{N_{\text{interval}}} \\
    &= \frac{1}{ N_{\text{interval}}}+ p Q_\omega^\text{Eve} \alpha + (1 - \beta) [\gamma Q_{\omega}^{\text{pass}} + (1 - \gamma) Y_0],\\
\end{aligned}
\end{equation}
}
where $N_\text{control}/N_\text{blind}/N_\text{dead}/N_\text{interval}$ is the gate number of the fully controllable range/the blinded period/the dead time/the interval length for a group of blinding pulses. $\alpha = N_{\text{control}}/N_{\text{interval}}$, representing the controllable proportion of gates under the attack. $1 - \beta = (N_{\text{interval}} - N_{\text{blind}} - N_{\text{dead}})/N_{\text{interval}}$, representing the proportion of gates that are not affected by the blinding. $p \in [0, 1]$ is the proportion of $N_\text{control}$ that Eve launches the fake-state attack. $\gamma \in [0, 1]$ is the ratio that Eve {\color{black} passes} the photons from Alice during the unblinded time in each interval. Accordingly, the total QBER is
{\color{black}
\begin{equation} \label{E}
\begin{aligned}
E_\omega &= \frac{1}{Q_\omega} \{\frac{e_0}{N_{\text{interval}}} + p Q_\omega^\text{Eve} \alpha e_\text{det} \\
    &+ (1-\beta)[\gamma E_{\omega}^{\text{pass}} Q_{\omega}^{\text{pass}} + (1 - \gamma) Y_0 e_0]\},
\end{aligned}
\end{equation}
}
where $E_\omega^{\text{pass}} Q_\omega^{\text{pass}} = e_0 Y_0 + e_\text{det}(1 - e^{- \eta_\text{Bob} \omega}).$ $e_0 = 0.5$ is the error rate of the background noise. $e_{\text{det}} = 3.3\%$ is the misalignment error rate of the QKD optical system.

According to the principle of the attack, Eve has to keep the total gain being indistinguishable with that in the normal working state~({\color{black}$Q_{\omega}^{\text{normal}} = Y_0 + 1 - e^{-\eta_\text{Bob} \eta_\text{ch} \omega}$, where $\eta_\text{ch} = 10^{\frac{-0.21 L}{10}}$} is the transmittance of the quantum channel of the QKD system as a function of the channel length $L$) to hide her existence by modulating $p$ and $\gamma$. Moreover, she will make {\color{black} $p$ ($\gamma$) as high (low) as possible}. Consequently, when Eve tries to make $p = 1$ and {\color{black}$\gamma = 0$} initially, she may confront with two cases:
\begin{itemize}
	\item [\emph{I:}] $Q_\mu > Q_\mu^{\text{normal}}$.
		In this case, Eve only needs to decrease $p$ to apply less fake-state attack during the fully controllable range to ensure $Q_\mu = Q_\mu^{\text{normal}}$. Thus, she can obtain almost all the information as all rounds of communication are either controlled or blocked.
	\item [\emph{II:}] $Q_\mu < Q_\mu^{\text{normal}}$.
		In this case, Eve has to {\color{black} increase} $\gamma$ to allow some photons pass from Alice to Bob without any intervention during the unblinded time, while keeps $p=1$, and then increase $Q_\mu$ to hide herself. Therefore, she can just obtain part of the total information in the communication.
\end{itemize}
Under this strategy, the QKD system cannot be aware of Eve's attack by checking the total gain. The QBER during the attack is shown in~\cref{fig:simulation}a.

\subsection{The key rate estimated by Alice and Bob under pulse illumination attack}
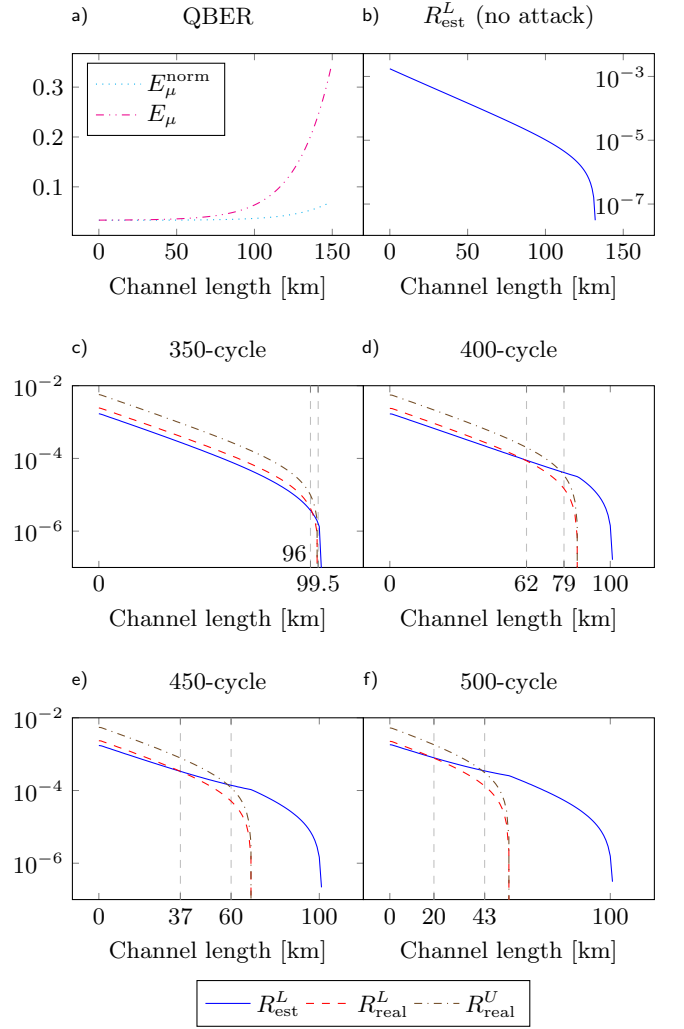
\begin{figure}
	\centering
  \begin{tikzpicture}[trim left=-0.8cm]
\begin{groupplot}[
group style={
	group name=my plots,
	group size=2 by 3,
	vertical sep=2cm,
	horizontal sep=0cm
},
ymode=log,
xmax=170,
width=5.45cm,
height=4cm,
extra x tick style={grid={major},grid style={dashed}, ticklabel style={anchor=south}},
extra x tick labels=\empty,
xlabel=Channel length,
x unit=\si{\kilo \meter},
major tick length=2.5pt,
]
\nextgroupplot [title=QBER, no markers, ymode=normal, legend entries={$E_{\mu}^{\text{norm}}$, $E_{\mu}$}, legend style={at={(0.05,0.95)}, anchor={north west}, cells={anchor=west}}]
\addplot+[cyan, dotted] table[col sep=comma, x=L, y=E_normal]  {data/simulation-data/E_norm.csv};
\addplot+[magenta, dashdotdotted] table[col sep=comma, x=L, y=E_attack]  {data/simulation-data/E_norm.csv};

\nextgroupplot [title=$R^L_\text{est}$ (no attack), yticklabel pos=right,yticklabel style={anchor=east},no markers]
\addplot+[] table[col sep=comma]  {data/simulation-data/RL_noattack.csv};

\nextgroupplot [title=350-cycle, ymin=1e-7, ymax=1e-2, no markers, extra x ticks={96, 99.5}, xtick={0, 99.5, 150}, xmax=120]
\addplot+[] table[col sep=comma, x=L, y=RL_esti]  {data/simulation-data/R_350.csv};
\addplot+[dashed] table[col sep=comma, x=L, y=RL_real]  {data/simulation-data/R_350.csv};
\addplot+[dashdotted] table[col sep=comma, x=L, y=RU_real]  {data/simulation-data/R_350.csv};
\node [xshift=2.6cm, yshift=-3.2cm] {96};

\nextgroupplot [title=400-cycle, ymin=1e-7, ymax=1e-2,ytick=\empty, no markers, extra x ticks={62, 79}, xtick={0, 62, 79, 100, 150}, xmax=120]
\addplot+[] table[col sep=comma, x=L, y=RL_esti]  {data/simulation-data/R_400.csv};
\addplot+[dashed] table[col sep=comma, x=L, y=RL_real]  {data/simulation-data/R_400.csv};
\addplot+[dashdotted] table[col sep=comma, x=L, y=RU_real]  {data/simulation-data/R_400.csv};

\nextgroupplot [title=450-cycle, ymin=1e-7, ymax=1e-2,no markers, extra x ticks={37, 60}, xtick={0, 37, 60, 100, 150}, xmax=120]
\addplot+[] table[col sep=comma, x=L, y=RL_esti]  {data/simulation-data/R_450.csv};
\addplot+[dashed] table[col sep=comma, x=L, y=RL_real]  {data/simulation-data/R_450.csv};
\addplot+[dashdotted] table[col sep=comma, x=L, y=RU_real]  {data/simulation-data/R_450.csv};

\nextgroupplot [title=500-cycle, ymin=1e-7, ymax=1e-2,ytick=\empty, no markers, legend columns=3, legend entries={$R^L_{\text{est}}$,$R^L_{\text{real}}$,$R^U_{\text{real}}$}, legend style={at={(0, -0.7)}, anchor={south}}, extra x ticks={20, 43}, xtick={0, 20, 43, 100, 150}, xmax=120]
\addplot+[] table[col sep=comma, x=L, y=RL_esti]  {data/simulation-data/R_500.csv};
\addplot+[dashed] table[col sep=comma, x=L, y=RL_real]  {data/simulation-data/R_500.csv};
\addplot+[dashdotted] table[col sep=comma, x=L, y=RU_real]  {data/simulation-data/R_500.csv};

\end{groupplot}
\draw [font=\sf \scriptsize](my plots c1r1.north west) node [xshift=0.1cm, yshift=0.5cm] {a)};
\draw [font=\sf \scriptsize](my plots c2r1.north west) node [xshift=0.1cm, yshift=0.5cm] {b)};
\draw [font=\sf \scriptsize](my plots c1r2.north west) node [xshift=0.1cm, yshift=0.5cm] {c)};
\draw [font=\sf \scriptsize](my plots c2r2.north west) node [xshift=0.1cm, yshift=0.5cm] {d)};
\draw [font=\sf \scriptsize](my plots c1r3.north west) node [xshift=0.1cm, yshift=0.5cm] {e)};
\draw [font=\sf \scriptsize](my plots c2r3.north west) node [xshift=0.1cm, yshift=0.5cm] {f)};
\end{tikzpicture}

  \caption{The simulation results of the security analysis. a) The QBER with/without Eve's pulse illumination attack. Here 350-/400-/450-/500-cycle attack introduce the same QBER. b) The $R^L_{\text{est}}$ when the system works without pulse illumination attack. The key rate decreases dramatically to almost 0 when the length is longer than \SI{130}{\kilo \meter}. c)/d)/e)/f) The $R^L_{\text{est}}$, $R^L_{\text{real}}$ and $R^U_{\text{real}}$ under Eve's pulse illumination attack with 350-/400-/450-/500-cycle illumination pulses.}
  \label{fig:simulation}
\end{figure}

According to the decoy-state protocol~\cite{ma2005, wang2005a}, Alice and Bob can estimate the yield and the error rate of a single photon, which are given by
\begin{equation}\label{YE_1}
\begin{aligned}
Y_1^L& =\frac{\mu}{\mu \nu-\nu^{2}}\left(Q_{\nu} e^{\nu}-Q_{\mu} e^{\mu} \frac{\nu^{2}}{\mu^{2}}-\frac{\mu^{2}-\nu^{2}}{\mu^{2}} Y_{0}\right)\\
e_1^U& =\frac{E_{\nu} Q_{\nu} e^{\nu}-e_{0} Y_{0}}{Y_{1}^{L} \nu}.
\end{aligned}
\end{equation}
Submitting \cref{Q}, \cref{E}, and \cref{YE_1} into the GLLP~\cite{gottesman2004}, Alice and Bob can estimate the lower bound of the key rate as
\begin{equation}\label{GLLP}
	R^L_{\text{est}} = q\left\{-Q_{\mu} f\left(E_{\mu}\right) H_{2}\left(E_{\mu}\right)+ \mu e^{-\mu} Y_{1}^L\left[1-H_{2}\left(e_{1}^U\right)\right]\right\}.
\end{equation}
Here $q=1/2$ for the BB84 protocol, $f(E_\mu)=1.2$ for error correction, and $H_2(x)$ is Shannon entropy. The $R^L_{\text{est}}$ under no/350-/400-/450-/500-cycle pulse illumination attack with Eve's strategy is shown by the blue line in~\cref{fig:simulation}b/c/d/e/f.

\subsection{The real key rate of the QKD system under pulse illumination attack}
To judge whether Alice and Bob overestimate the key rate and thus introduce insecurity, we give the \emph{real} upper bound and the lower bound of the key rate when the pulse illumination attack with Eve's strategy is applied.

The real yield of a single photon $Y_1^{\text{attack}}$ and its error rate $e_1^{\text{attack}}$ under Eve's attack strategy can be calculated as
\begin{equation}
\begin{aligned}
	Y_1^{\text{attack}} &= Y_0 +\eta_{\text{Bob}} -Y_0 \eta_{\text{Bob}},\\
	e_1^{\text{attack}} &=\frac{1}{Y_1^{\text{attack}}} (e_{\text{det}} \eta_{\text{Bob}} + e_0Y_0).
\end{aligned}
\end{equation}
Then, the real upper bound and the lower bound of the key rate can be written as
\begin{equation}\label{True_kr_upper}
	R^U_{\text{real}} = \frac{1}{2} (1-\beta)\gamma \mu e^{-\mu} Y_1^{\text{attack}}  [1-H_2(e_1^{\text{attack}})]
\end{equation}
and
\begin{equation}\label{True_kr_lower}
\begin{aligned}
	R^L_{\text{real}} &= \frac{1}{2} (1-\beta)\gamma \{\mu e^{-\mu} Y_1^{\text{attack}}  [1-H_2(e_1^{\text{attack}})] \\
	&- Q_\omega^{\text{pass}} f_{EC} H_2(E_\omega^{\text{pass}})\}.
\end{aligned}
\end{equation}
$R^L_{\text{real}}$ and $R^U_{\text{real}}$ under 350-/400-/450-/500-cycle pulse illumination attack with Eve's strategy are shown in~\cref{fig:simulation}c/d/e/f as the dashed and dash-dot lines.
The results show that Eve can successfully hack the QKD system and learn the secret key under certain communication distance between Alice and Bob for the cases considered here. Take the scenario of 500-cycle pulse illumination attack as an example, we can easily find that when the channel length is between 20 km and 43 km, the estimated key rate by Alice and Bob is higher than the real lower bound but lower than the real upper bound. Thus, Alice and Bob overestimates the key rate. When the length of the quantum channel is longer than \SI{43}{\kilo \meter}, we can ensure that the key rate estimated by Alice and Bob is insecure, because $R^L_{\text{est}}$ is higher than $R^U_{\text{real}}$.
{\color{black}
These results are reasonable.
When the channel length is short Eve has to pass a large portion of signals from Alice, and thus her threat to the QKD system is weak.
The GLLP equation still can estimate a secret key rate in the secure range.
However, as the channel length gets longer, Eve can block more signals during the unblinded time to enlarge the proportion of the eavesdropped keys, and thus threaten the security significantly.
The GLLP equation then cannot correctly estimate the secret key rate.
}
\section{Discussion}
\label{sec:disc}
As we described above, the pulse illumination attack can hack the passive quenching single photon detector to eavesdrop secret information while bypass its photocurrent monitor.
Here we extend the discussion to the effectiveness of the pulse illumination attack on detectors of other architectures or with other countermeasures.
The pulse illumination attack can blind high-speed self-differential detectors~\cite{dynes2016, yuan2018}, as this kind of detectors can be blinded by triggering a sequence of detection events~\cite{jiang2013}, which can be achieved by the pulse illumination attack with a low blinding pulse energy.
For active-quenching detectors without a bias resistor~\cite{zheng2018}, blinding by bright illumination seems difficult.
However, the work of thermal attack shows that this kind of detectors can also be blinded by the thermal effect of the injected bright light~\cite{lydersen2010b}, which implies that the pulse illumination attack with a higher blinding pulse intensity may hack such detectors.
{\color{black} In this case, the pulse illumination attack tends to introduce a relatively higher generated photocurrent which might reveal the existence of Eve.
However, Eve can further enlarge the inter-group interval to decrease the low-frequency components of the generated photocurrent and hide herself again.}
A sophisticated attack-monitoring method shown in Ref.~\cite{silva2012} exploits the accumulating statistics of times between consecutive detection events.
As the pulse illumination attack has various adjustable parameters, its fingerprint on the statistics might be attenuated.
Recently, a countermeasure against detector-control attacks using randomly switching variable attenuators (VA) is proposed~\cite{qian2019}, where the switching of a VA's attenuation value  under c.w.\ illumination will introduce random clicks and raise QBER to trigger the alarm.
However, the pulse illumination attack does not illuminate constantly, and thus the switching probably does not lead to any abrupt changes on injected light to cause random clicks.
The practical effectiveness of these countermeasures under the pulse illumination attack should be analysed in a future testing.

{\color{black} 
Lowering the alarming threshold of the photocurrent monitor can not effectively detect the pulse illumination attack.
In addition to frequent false alarms introduced by this method, Eve can attenuate the reported photocurrent by enlarging the intervals to break this defense, as shown in \cref{sec:detail}.
Improving the stop band of the filter in the photocurrent monitor can reveal more evidences of the pulse illumination attack, depending on the value improved.
However, improving the stop band extremely will also make the monitor in trouble with abundant false alarms by noises and lose its functionality.
An extreme case is removing the filter to thoroughly expose the uprising caused by the blinding pulses.
In this case, the alarm will definitely be triggered with the pulse illumination attack, but frequent false alarms will also be trigger when the detector is working normally.
}
The feasibility of {\color{black}any change of parameters in the photocurent monitor} should be tested in a future experiment, and the change itself might introduce some new loopholes.
To patch the loophole exploited by the pulse illumination attack and defend this kind of blinding attacks thoroughly, we believe that the {\color{black}valid} method is to integrate the loophole into the security proof of QKD protocol.

\section{Conclusion}
We investigate the effectiveness of a photocurrent monitor as a countermeasure against the detector blinding attack in a single-photon detector module that is provided by an independent party. The testing results show that the single-photon detector with a photocurrent monitor is vulnerable to the pulse illumination attack. Via this attack, Eve can blind the single-photon detector in a certain period and fully control its detection output, keeping the reported photocurrent of the photocurrent monitor similar to that in the normal state and thus without alarming the monitor. We also perform the theoretical security analysis to show that for a real-life QKD system under pulse illumination attack, Alice and Bob may overestimate the secret key rate and leak the key to Eve in a certain distance range. This pulse illumination attack indicates that the security issues in the detection side {\color{black} might be} still serious, which should be further investigated. As this attack {\color{black}might} seriously threatens the practical security of QKD systems, pulse illumination attack should be a standard testing item for the systematic security evaluation of a QKD system.

We also provide more details on the photocurrent of a detector under the pulse illumination attack obtained from a white-box experiment on our homemade detector (see \cref{sec:countermeasure}), which may provide some ideas of countermeasures against the pulse illumination attack. However, patching only solves the problem in a short term. A more secure method is to model the practical single-photon detector in the security proof, if the non-MDI-QKD system would like to be immune to various blinding attacks in a long term.

\section*{Funding Information}
National Natural Science Foundation of China (Grants No. 61901483, No. 11674397, No. 61601476, and No. 61632021) and National Key Research and Development Program of China (Grant No. 2019QY0702).

\section*{Acknowledgments}

We thank Vadim Makarov for very useful discussions. Supporting from Greatwall Quantum Laboratory is also acknowledged.

\section*{Disclosures}

\noindent\textbf{Disclosures.} The authors declare no conflicts of interest.




\appendix

\section*{Appendix}

\section{Recap c.w. illumination blinding attack}
\label{sec:c.w.}

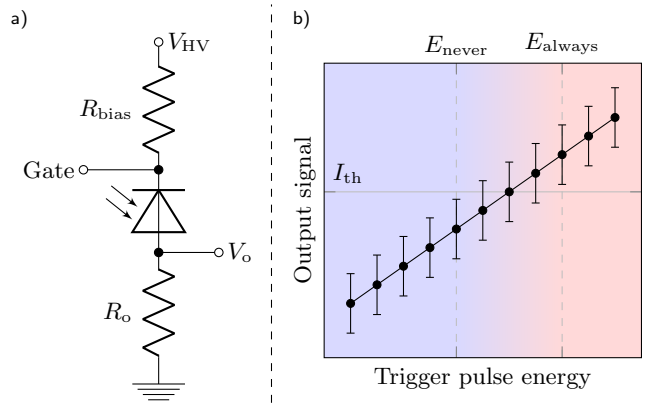
\begin{figure}\centering
  \begin{tikzpicture}

\begin{axis} [
at={(17mm,-5mm)},
set layers,
width=5.8cm,
height=5.5cm,
xtick=\empty,
ytick=\empty,
xlabel={Trigger pulse energy},
xlabel shift={-5mm},
ylabel={Output signal},
extra x ticks={0.5, 0.9},
extra x tick labels={$E_{\text{never}}$, $E_{\text{always}}$},
extra x tick style={grid={major},grid style={dashed},xticklabel pos=top},
extra y ticks={3.5},
extra y tick labels={$I_{\text{th}}$},
extra y tick style={grid=major,ticklabel style={anchor=south west}},
]
\addplot+ [
black,
mark color=black,
mark size=1.5pt,
mark options={
fill=black,draw=black,
},
error bars/.cd,
y dir=both,y fixed =0.8,
] coordinates {
(0.1,0.5)
(0.2,1)
(0.3,1.5)
(0.4, 2)
(0.5, 2.5)
(0.6, 3)
(0.7, 3.5)
(0.8, 4)
(0.9, 4.5)
(1, 5)
(1.1, 5.5)
};
\pgfonlayer{axis background}
\fill [left color=blue!15, right color=red!15,on layer=axis background] (rel axis cs:0.42, 0) rectangle (rel axis cs:0.75,1);
\fill [color=blue!15,on layer=axis background] (rel axis cs:0,1) rectangle (rel axis cs:0.42, 0);
\fill [color=red!15,on layer=axis background] (rel axis cs:0.75, 0) rectangle (rel axis cs:1,1);
\endpgfonlayer

\end{axis}
\draw [xshift=-0.5cm, yshift=-0.5cm] (0,0) node[ground] {} to [R, l=$R_{\text{o}}$] (0,12mm) to (0,15mm) to[stroke photodiode] (0,24mm) to [R, l=$R_{\text{bias}}$, -o] (0mm,42mm) node[anchor=west] {$V_{\text{HV}}$};

\draw [xshift=-0.5cm, yshift=-0.5cm] (0,14mm) to[short, *-o] (8mm,14mm) node[anchor=west] {$V_{\text{o}}$} (0, 25mm) to[short, *-o] (-10mm, 25mm) node[anchor=east] {Gate};
\node [font=\sf \scriptsize, at={(-23.5mm,40mm)}] {a)};
\node [font=\sf \scriptsize, at={(14mm,40mm)}] {b)};
\draw [dashed](10mm,-11mm) to (10mm, 42mm);
\end{tikzpicture}

  \caption{Inner mechanism of the single-photon detector. a) The core part of the circuit of a typical single-photon detector. $R_{\text{bias}}$ is a huge resistor for passive quenching while $R_{\text{o}}$ is a small resistor for readout. The voltage across $R_{\text{o}}$ is $V_{\text{o}}$, which is the carrier of the output signals. $V_{\text{HV}}$ is the DC source of the single-photon detector's circuit. $V_{\text{bias}}$ is the bias voltage across the APD. Normally, $V_{\text{bias}}$ is lower than the breakdown voltage~($V_{\text{br}}$) and can be raised to be higher than it by gate signals. b) Schematic diagram of the relationship between the trigger pulse energy and the responding output signal when the APD is in the linear mode. $I_{\text{th}}$ is the threshold of a built-in comparator in the circuit of the single-photon detector. $E_{\text{never}}$/$E_{\text{always}}$ is the trigger pulse's energy that triggers a click with 0\%/100\% probability, because the bound of its error bar is totally lower/higher than $I_{\text{th}}$. Two different background colors intuitively indicate whether the trigger pulse can trigger a click.}
  \label{fig:apd-inner}
\end{figure}

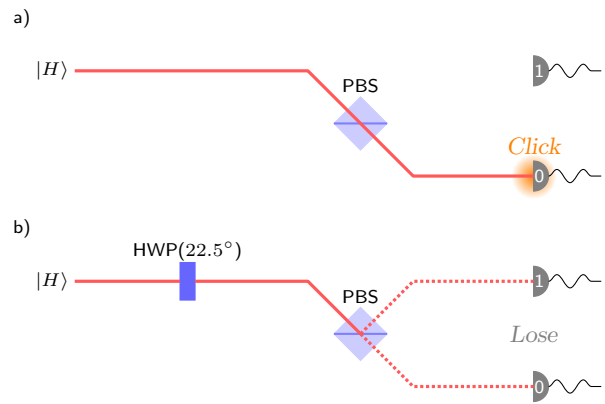
\begin{figure}\centering
  \begin{tikzpicture}
[headers/.style={font=\sf \scriptsize}]
\def \PBS#1{
    \filldraw [blue!20, rotate=-45] (-2.5mm,-2.5mm) rectangle ++(5mm,5mm);
    \draw [rotate=-45, thick, blue!50] (-2.5mm, -2.5mm) -- (2.5mm, 2.5mm);
    \node [headers]at (0, 5mm) {PBS};
    \coordinate (#1_lt) at (-7mm, 7mm);
    \coordinate (#1_lb) at (-7mm, -7mm);
    \coordinate (#1_c) at (0mm, 0mm);
    \coordinate (#1_rt) at (7mm, 7mm);
    \coordinate (#1_rb) at (7mm, -7mm);
}
\def \HWP#1{
    \filldraw [blue!60] (-1mm,-2.5mm) rectangle (1mm,2.5mm);
    \node [headers]at (0, 4mm) {HWP($22.5^\circ$)};
    \coordinate (#1_l) at (-1mm, 0mm);
    \coordinate (#1_r) at (1mm, 0mm);
}
\def \APDZ#1{
	\filldraw  [black!50] (0mm,-2mm) arc [start angle=-90, end angle=90, radius=2mm] -- cycle;
	\draw [decorate, decoration=snake] (2mm, 0) -- (9mm, 0);
	\node [headers, white] at (0.8mm, 0) {0};
    \coordinate (#1_l) at (0mm, 0mm);
}
\def \APDO#1{
	\filldraw  [black!50] (0mm,-2mm) arc [start angle=-90, end angle=90, radius=2mm] -- cycle;
	\draw [decorate, decoration=snake] (2mm, 0) -- (9mm, 0);
	\node [headers, white] at (0.8mm, 0) {1};
    \coordinate (#1_l) at (0mm, 0mm);
}
\def \SrcH#1{
    \node [headers] at (-3mm, 0) {$\Ket{H}$};
    \coordinate (#1_c) at (0mm, 0mm);
}
\def \Click#1 {
    \shade [inner color=orange, radius=3mm] (0, 0) circle; 
    \node [font=\slshape \color{orange}] at (0, 4mm) {Click};
}
\def \Lose#1 {
    \node [font=\slshape \color{gray}] at (0, 0) {Lose};
}
\matrix (m) at (0, 0)
[
    row sep={7 mm,between origins},
    column sep={23 mm,between origins}
]
{
    \node[headers, xshift=-7mm] at (0, 0) {a)}; &[-8mm] & & \\
    \SrcH{srch_hv}& & & \APDO{apd1_hv} \\
    & & \PBS{pbs_hv}& \\
    & & & \Click{click_hv} \APDZ{apd0_hv} \\
    \node[headers, xshift=-7mm] at (0, 0) {b)};& & & \\
    \SrcH{srch_fs}& \HWP{hwp_fs}& & \APDO{apd1_fs} \\
    & & \PBS{pbs_fs}& \Lose{lose_fs} \\
    & & & \APDZ{apd0_fs} \\
};
\begin{scope}[red!65, very thick, line join=round]
\draw (srch_hv_c) -- (pbs_hv_lt) -- (pbs_hv_c) -- (pbs_hv_rb) -- (apd0_hv_l);
\draw (srch_fs_c) -- (hwp_fs_l);
\draw (hwp_fs_r) -- (pbs_fs_lt) --  (pbs_fs_c);
\end{scope}
\begin{scope}[red!65, very thick, densely dotted, line join=round]
\draw (pbs_fs_c) -- (pbs_fs_rt) -- (apd1_fs_l);
\draw (pbs_fs_c) -- (pbs_fs_rb) -- (apd0_fs_l);
\end{scope}
\end{tikzpicture}

  \caption{Faked-state attack on the blinded detectors when Eve chooses the $H$/$V$ basis and the intercepted measurement result is $\ket{H}$. Eve resends a trigger pulse of $\ket{H}$ in $[E_{\text{always}}, 2 \times E_{\text{never}})$. Only the situation that Bob chooses matching basis with Alice is discussed here. a) Bob selects the same basis ($H$/$V$) with Eve. Subsequently, the full trigger pulse transmits through the polarizing beam splitter (PBS) and triggers a click that means 0, which is identical with Eve's measurement result. b) Bob selects the opposite basis ($+$/$-$). Half energy of the trigger pulse, which is less than $E_{\text{never}}$, arrives at each single-photon detector as the dashed line shows, and none of them is triggered. In a word, Eve steals an effective bit when she chooses the matching basis, while blocks the bit as her basis mismatches.}
  \label{fig:msrsetup}
\end{figure}


The BB84 protocol is typically used in QKD implementation, especially in commercial QKD systems. Under this protocol, a QKD system can provide a information-theoretical secure communication channel to the legitimate users, as the existing of an eavesdropper will introduce 25\% or higher QBER and trigger the alarm~\cite{mayers1996, lo1999, shor2000}.

Note that the above scenario only works well under the assumption that the APDs works in Geiger mode~\cite{cova2004}, where a single photon leads to huge transient avalanche photocurrent and thus causes a click. However, a real-life QKD system deviates from the ideal model: the APD can be turned into the linear mode~(be blinded) and then the clicks are controlled by Eve. One approach to achieve blinding is to illuminate the APD by carefully modulated c.w.\ light. The principle behind the c.w.\ illumination blinding is as follows. Bright c.w.\ light applied on the APD knocks out many electron-hole pairs, and thus a huge photocurrent is generated. According to the circuit shown in~\cref{fig:apd-inner}a, the APD is in series with the $R_{\text{bias}}$, so the strong photocurrent also goes through the $R_{\text{bias}}$. Because the $R_{\text{bias}}$ is a huge resistor for passive quenching, the voltage across $R_{\text{bias}}$ increases dramatically and thus $V_{\text{bias}}$ goes lower than $V_{\text{br}}$ as the total voltage is conserved.

Having been blinded, the single-photon detectors of Bob can be controlled secretly by applying trigger pulses. A trigger pulse can always trigger a click when its energy is higher than $E_{\text{always}}$, and is impossible to trigger a click when the energy is lower than $E_{\text{never}}$, as shown in the~\cref{fig:apd-inner}b. More specifically, with the assumptions that the all of Bob's single-photon detectors are identical and satisfy
\begin{equation} \label{eq:controll sufficient}
    E_{\text{always}} < 2 \times E_{\text{never}},
\end{equation}
Bob can be fully controlled by the fake-state attack \cite{makarov2005} if the energy of the trigger pulse in $[E_{\text{always}}, 2 \times E_{\text{never}})$. In a round of the communication Eve intercepts the photon and measures it in a randomly chosen basis, and then resends a trigger pulse encoded by the measurement result to Bob. Eve can ensure that when she happens to choose the same basis as both Alice and Bob, the information of this round will be shared among Alice, Bob, and Eve~(see~\cref{fig:msrsetup}a), while Bob will get no click and lose the information when Eve unfortunately chooses a wrong basis~(see~\cref{fig:msrsetup}b). Afterwards, with the help of basis comparison in the post-processing~(which happens in a public classical channel and can be listened by Eve), all of Alice, Bob, and Eve keep the bits that are measured with same basis. As a result, Alice and Bob innocently share the final entire secret key with Eve.


\section{Detailed analysis on the parameters of blinding pulses}
\label{sec:detail}
\begin{figure}\centering
  \begin{tikzpicture}
\begin{axis}
[
    set layers,
    x SI prefix=pico, x unit=\si{\joule},
    y unit=\si{\micro \ampere},
    legend pos=south east,
    legend entries={\SI{500}{\nano \second},\SI{600}{\nano \second},\SI{700}{\nano \second}},
    extra y ticks={31},
    extra y tick style={grid={major},grid style={dashed}, ticklabel style={anchor=south west}},
    xlabel={Energy of single blinding pulse},
    ylabel={Reported photocurrent},
]
\addplot+ table[col sep=comma, x=E, y=I]  {data/EvesusI_500ns.csv};
\addplot+ table[col sep=comma, x=E, y=I]  {data/EvesusI_600ns.csv};
\addplot+[mark=triangle*] table[col sep=comma, x=E, y=I]  {data/EvesusI_700ns.csv};
\pgfonlayer{axis foreground}
\node [align=center] at (axis cs:1,34.5) {Constant blinded \\ threshold};
\draw [-stealth,thick, orange] (axis cs: 2.5175,32.4) -- (axis cs:2.5175,26.9);
\draw [-stealth,thick, blue] (axis cs: 2.5175,32.4) -- (axis cs:3.5817,32.4);
\endpgfonlayer
\end{axis}
\end{tikzpicture}

  \caption{The reported photocurrent versus the energy of single blinding pulse with interval of \SI{500}{\nano \second}/\SI{600}{\nano \second}/\SI{700}{\nano \second}. Here the blinding pulses are in the simplest case, that is, 1-cycle per group. Reported photocurrent higher than the constant blinded threshold~(\SI{31}{\micro \ampere}) indicates the APD is blinded in whole time domain.}
  \label{fig:EvesusI}
\end{figure}
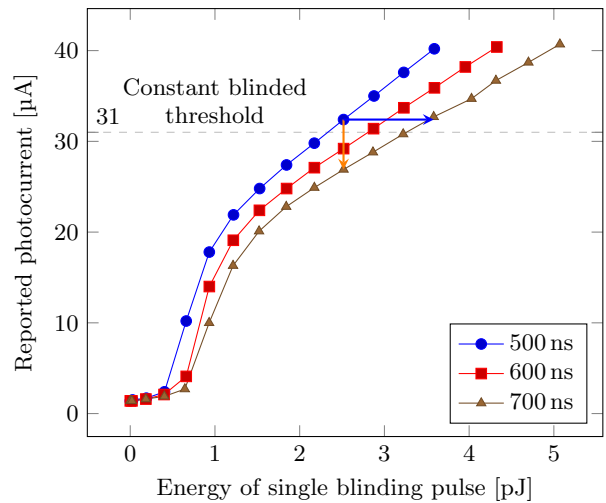
Here we first demonstrate the simplest case -- 1-cycle blinding pulses in each group. In the experiment, we controlled the interval between each group of blinding pulse and the energy of each single pulse. Then we observed the reported photocurrent. \cref{fig:EvesusI} shows the energy of each single blinding pulses versus the reported photocurrent with interval of 500ns/600ns/700ns. Generally, the reported photocurrent increases with the rising of single pulse energy. The reported photocurrent rises slightly at the beginning and then goes up dramatically at about \SI{0.67}{\pico \joule}. Finally, the reported photocurrent ascends linearly after \SI{0.9}{\pico \joule}. In addition, in~\cref{fig:EvesusI}, the points where the reported photocurrent is higher than \SI{31}{\micro \ampere} means that the low-frequency component is strong enough to blind the APD in the whole time domain (\emph{constant blinding}).

The orange vertical arrow in \cref{fig:EvesusI} shows that the reported photocurrent reduces as the interval rises. This is because for the same energy of blinding pulse, the larger interval between the groups results in the less generated photocurrent from different blinding pulses that superposes with each other. Consequently, the low-frequency components of the superposed photocurrent are less, which are reported by the photocurrent monitor. Contrarily, to increase the superposed photocurrent to constantly blind the APD when the interval is extended, higher energy of each blinding pulse is needed, as shown in the blue arrow in~\cref{fig:EvesusI}. From the testing result, we can see that Eve can extend the interval to reduce the reported photocurrent, and thus avoiding the alarm of the photocurrent monitor.

\begin{figure}\centering
  \begin{tikzpicture}
\begin{axis}
[
    xlabel={Interval length},
    ylabel={Constant blinding energy},
    x unit=\si{\micro \second},
    y SI prefix=pico, y unit=\si{\joule},
    legend entries={1-cycle,2-cycle,3-cycle},
    legend pos=south east,
]
\addplot+ [] table[col sep=comma, x=T, y=Eperintrv]  {data/1cyclesplit.csv};
\addplot+ [] table[col sep=comma, x=T, y=Eperintrv]  {data/2cyclesplit.csv};
\addplot+ [mark=triangle*] table[col sep=comma, x=T, y=Eperintrv]  {data/3cyclesplit.csv};
\end{axis}
\end{tikzpicture}

  \caption{Constant-blinding energy versus interval length of the blinding pulses. The constant~blinding energy is defined as the total energy of each group with 1/2/3-cycle blinding pulses to blind the APD in the whole time domain.}
  \label{fig:cyclesplit}
\end{figure}
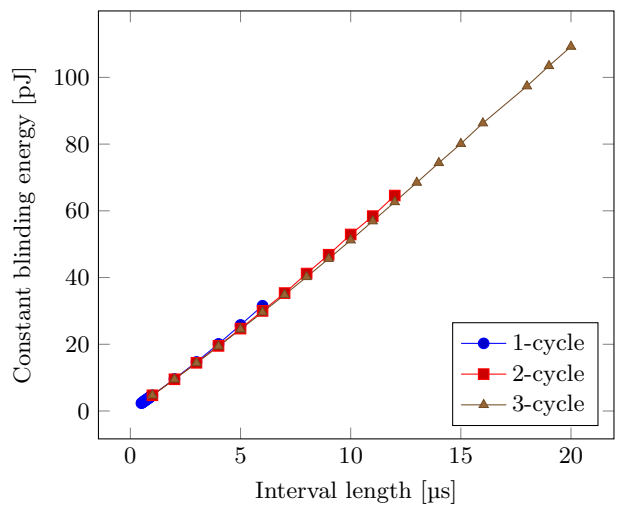

To further analyse the influence introduced by the cycle number in each group, we measured the total energy of each group with 1-/2-/3-cycle blinding pulses when the APD is constantly blinded in the whole time domain~(which is defined as \emph{constant-blinding energy} in the following text) versus the interval length. The measurement results are shown in~\cref{fig:cyclesplit}. Comparing among the three curves in~\cref{fig:cyclesplit}, for the same interval length, the summation energies of each 1-/2-/3-cycle group to constantly blind the APD are quite similar. Thus the equivalence between a single blinding pulse and three smaller blinding pulses is apparent. Moreover, the maximum intervals for the 1-/2-/3-cycle blinding pulses are \SI{6}{\micro \second}/\SI{12}{\micro \second}/\SI{20}{\micro \second} respectively. Taking the case of 1-cycle for illustration, if the interval is longer than its maximum value, the increased energy of pulses will no longer blind the APD but cause unwanted clicks. However, its equivalent split in 2-/3-cycle can still blind. Therefore, by using this multi-cycle approach, the corresponding blinded period is adjustable in a wider range.

\section{The waveform of a homemade detector under pulse illumination attack}
\label{sec:countermeasure}
\begin{figure}\centering
  \begin{tikzpicture}
\begin{groupplot}[
group style={
group size=1 by 2,
x descriptions at=edge bottom,
vertical sep=0.7cm
} ,
change y base = true,
width=8cm,
height=5cm,
xlabel=Time,
x SI prefix=nano, x unit=\si{\second},
y SI prefix=milli, y unit=\si{\volt},
ymax=0.025,
ymin=-0.028,
clip=false
]

\nextgroupplot [ylabel=Without blinding]
\addplot[cyan] table[col sep=comma] {data/homade-APD-data/b11.csv};
\addplot[cyan] table[col sep=comma] {data/homade-APD-data/b12.csv};
\addplot[cyan] table[col sep=comma] {data/homade-APD-data/b13.csv};
\draw [dashed, gray] (axis cs: 25.9,0.0075) -- (axis cs:31.9,0.0075);
\draw [|<->|, thick] (axis cs: 28.9,0) -- (axis cs:28.9,0.012);
\node [xshift=-15mm, yshift=37mm,font=\sf \scriptsize] {a)};

\nextgroupplot [ylabel=After blinding]
\addplot[magenta] table[col sep=comma] {data/homade-APD-data/a15.csv};
\addplot[magenta] table[col sep=comma] {data/homade-APD-data/a16.csv};
\addplot[magenta] table[col sep=comma] {data/homade-APD-data/a17.csv};
\draw [dashed, gray] (axis cs: 25.9,0.0075) -- (axis cs:31.9,0.0075);
\draw [|<->|,thick] (axis cs: 28.59,0) -- (axis cs:28.59,0.0055);
\node [xshift=-15mm, yshift=37mm,font=\sf \scriptsize] {b)};

\end{groupplot}
\end{tikzpicture}

  \caption{The 10-sample overlaid waveform of $V_{\text{o}}$ of our homemade single-photon detector a) without/ b) after blinding pulses. The dashed line indicates the comparator threshold for triggering a click. Here the comparator only works on the timing when the output signals may occur. The fluctuation ranges of the output signals are indicated by the double-head arrows.}
  \label{fig:current-waveform}
\end{figure}
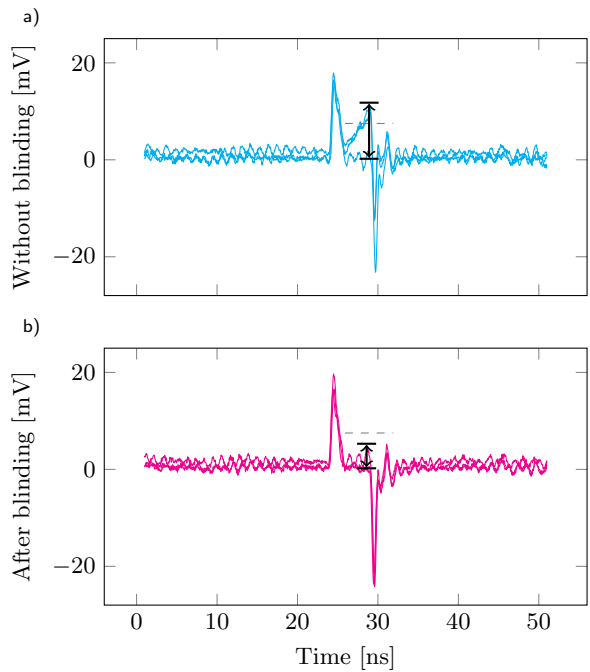
\begin{figure}\centering
  \begin{tikzpicture}
\begin{axis}
[
  change y base = true,
  xlabel={Time},
  ylabel={$V_{\text{o}}$},
  x SI prefix=nano, x unit=\si{\second},
  y SI prefix=milli, y unit=\si{\volt},
]
\addplot+ [no markers,magenta] table[col sep=comma] {data/homade-APD-data/pulseform.csv};

\end{axis}
\end{tikzpicture}

  \caption{The waveform of $V_\text{o}$ of our homemade single-photon detector when a blinding pulse arrives the APD. A huge instantaneous photocurrent hill is caused by the blinding pulse. The smaller one is caused by the gate signal.}
  \label{fig:pulseform}
\end{figure}
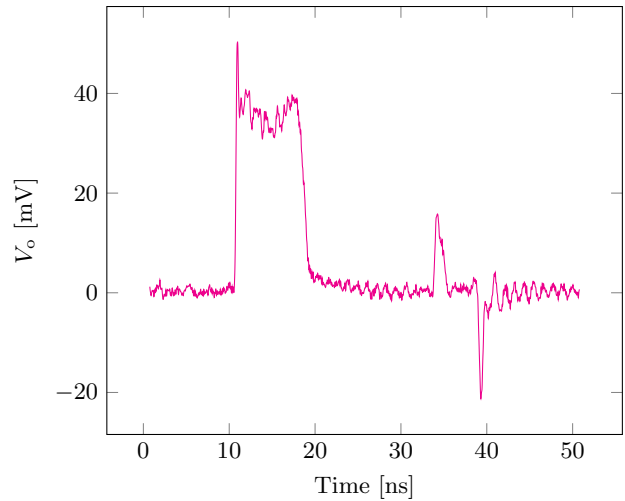
To acquire some evidences left by a pulse illumination attack, we conducted a white-box test of the pulse illumination attack on our homemade single-photon detector whose APD is produced by Princeton Lightwave. We directly observed the waveform of the voltage $V_{\text{o}}$ across the readout resistor $R_{\text{o}}$~(for our homemade single-photon detector, $R_\text{o} = \SI{50}{\ohm}$) in the circuit of the single-photon detector shown in~\cref{fig:apd-inner}. The waveform of $V_{\text{o}}$ without/after blinding pulses are compared in \cref{fig:current-waveform}. The fluctuation ranges of the output signals are indicated by the double-head arrows in the picture. Apparently, as shown in \cref{fig:current-waveform}a, when the blinding pulses are not applied, output signals rise occasionally. These rises are strong enough to reach the comparator threshold to trigger a click because of the avalanche effect in Geiger mode. As a result, dark counts are caused in this case. On the other hand, in the blinded period under the pulse illumination attack, the output signals jump frequently but just in a much narrower range whose upper bound is far lower than the comparator threshold as shown in  \cref{fig:current-waveform}b. Thus, dark counts are eliminated in the blinded period.

Another interesting evidence is shown in \cref{fig:pulseform}, where we can clearly see that a blinding pulse causes a huge instantaneous photocurrent hill. A photocurrent monitor might be capable of figuring out this evidence by some engineering modifications, which may be not easy to be realized. Researches on reliable countermeasures against pulse illumination attack are still urgently needed.

\bibliography{paper.bib}

\end{document}